\documentclass[aip,graphicx,reprint]{revtex4-1}
\usepackage{graphicx}
\usepackage{dcolumn}
\usepackage{bm}

\draft 

\begin{document}





\title{Combination of genetic crossover and replica-exchange method
for conformational search of protein systems} 

\author{Yoshitake Sakae}
\affiliation{Department of Physics, Graduate School of Science, Nagoya University, 
Nagoya, Aichi 464-8602, Japan}
\author{Tomoyuki Hiroyasu}
\affiliation{Department of Biomedical Information, Doshisha University, Kyotanabe, Kyoto 610-0394, Japan}
\author{Mitsunori Miki}
\affiliation{Department of Intelligent Information Engineering and Sciences, Doshisha University, Kyotanabe, Kyoto 610-0394, Japan}
\author{Katsuya Ishii}
\affiliation{Information Technology Center, Nagoya University, Nagoya, 
Aichi 464-8601, Japan}
\author{Yuko Okamoto}
\affiliation{Department of Physics, Graduate School of Science, Nagoya University, 
Nagoya, Aichi 464-8602, Japan}
\affiliation{Information Technology Center, Nagoya University, Nagoya, 
Aichi 464-8601, Japan}
\affiliation{Structural Biology Research Center, Graduate School of Science, Nagoya University, Nagoya, Aichi 464-8602, Japan}
\affiliation{Center for Computational Science, Graduate School of Engineering, Nagoya University, Nagoya, Aichi 464-8603, Japan}



















\begin{abstract}
We combined the genetic crossover, which 
is one of the operations of genetic algorithm, and replica-exchange method
in parallel molecular dynamics
simulations. 
The genetic crossover and replica-exchange method can search the global 
conformational space by exchanging the 
corresponding parts between a pair of conformations of a protein. In this study, 
we applied this method to an $\alpha$-helical protein, Trp-cage mini protein, which 
has 20 amino-acid residues. The conformations obtained from the simulations are 
in good agreement with the experimental results.
\end{abstract}


\maketitle 



\section{Introduction}
The search of the stable states of biomolecules, such as DNA and proteins, by molecular simulations is important to understand their functions and stabilities.
However, as the biomolecules have a lot of local minimum-energy states separated by high energy barriers,
conventional molecular dynamics (MD) and Monte Carlo (MC) simulations tend to get trapped in states of
local minima.
To overcome this difficulty,
various sampling and optimization methods for conformations of biomolecules have been proposed
such as generalized-ensemble algorithms\cite{GEA} which include
the multicanonical algorithm (MUCA) \cite{MUCA1,MUCA2}, simulated tempering (ST) \cite{ST1,ST2}
and replica-exchange method (REM) \cite{REM1,REMD}.
We have also proposed a conformational search method referred to as the 
parallel simulated annealing using genetic crossover 
(PSA/GAc) \cite{PSAGAc,PSAGAc2,PSAGAc3,PSAMDGAc},
which is a hybrid algorithm combining both simulated annealing (SA) \cite{SA}
and genetic algorithm (GA) \cite{GA1,GA2}.
In this method, parallel simulated annealing simulations are combined
with genetic crossover,
which is one of the operations of genetic algorithm.
Moreover, we proposed a method that combines parallel MD simulations
and genetic crossover with Metropolis criterion \cite{PMDGAc}.

In this study, we applied this latest conformational search method \cite{PMDGAc}
using the genetic crossover 
to Trp-cage mini protein, which has 20 residues.
The operation of the genetic crossover is combined with the conventional MD and REM.
The obtained conformations during the simulation are in good agreement with the 
experimental results. This article is organized as follows. In Section 2 we
explain the present methods. In Section 3 we present the results. Section 4
is devoted to conclusions.

\section{Methods}
\label{method}
\subsection{Parallel molecular dynamics using genetic crossover}
We briefly describe our method \cite{PMDGAc}.
We first prepare $M$ initial conformations of the system in study, 
where $M$ is the total number of ``individuals'' in genetic algorithm
and is usually taken to be an even integer.
We then alternately perform the following two steps:
\begin{enumerate}
\renewcommand{\labelenumi}{\arabic{enumi}.}
\item 
For the $M$ individuals, regular canonical MC or MD simulations 
at a fixed temperature $T$ are carried out simulataneously and independently 
for a certain MC or MD steps.
\item 
$M/2$ pairs of conformations are selected from ``parental'' 
group randomly, and the crossover and selection operations are performed.
Here, the parental group means the latest conformations obtained in Step 1.
\end{enumerate}
 
\begin{figure}
\begin{center}
\resizebox*{8cm}{!}{\includegraphics{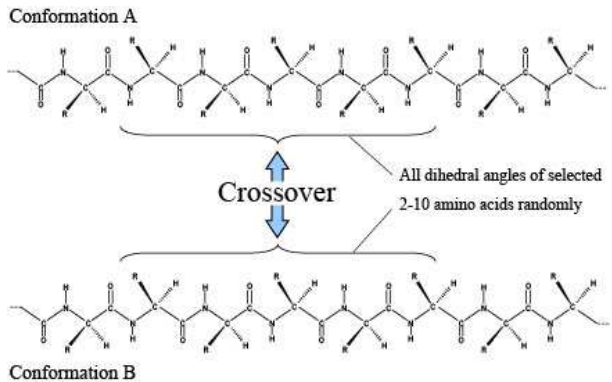}}%
\caption{Schematic process of the two-point crossover operation.
In this process, all dihedral angles (in backbone and side chains) within the randomly selected $n$ consecutive amino acids are exchanged 
between a pair of conformations.}
\label{fig_crossover}
\end{center}
\end{figure}

If we employ MC simulations in Step 1 above, we can refer the
method to as parallel Monte Carlo using genetic crossover (PMC/GAc)
and if MD simulations, parallel molecular dynamics using genetic 
crossover (PMD/GAc). 
In Step 2,
we can employ various kinds of genetic crossover operations.
Here, we just present a case of the 
two-point crossover (see Ref.~\cite{PSAMDGAc}).
The following procedure is carried out (see Fig.~\ref{fig_crossover}) :\\
%

\begin{enumerate}
\item Consecutive amino acids of length $n$ residues 
in the amino-acid sequence of the conformation are selected 
randomly for each pair of selected conformations.
\item Dihedral angles (in only backbone or all dihedral angles) in 
the selected $n$ amino acids are exchanged between the selected 
pair of conformations. 
\end{enumerate}
Note that the length $n$ of consecutive amino-acid residues can, 
in general, be different for each pair of selected conformations.
  
We need to deal with the produced ``child'' conformations with care.
Because the produced conformations often have unnatural structures 
by the crossover operation, they 
have high potential energy and are unstable.
Therefore, a relaxation process is introduced before the selection operation.
Short simulations at the same temperature $T$
with restraints 
on the backbone dihedral angles of only the $n$ 
amino acids are performed so that the corresponding backbone
structures of the $n$ amino acids 
will approach the exchanged backbone conformation.
The initial conformations for these equilibration simulations
are the ones before the exchanges.
Namely, by these equilibration simulations,
the corresponding backbone conformations of the $n$ amino acids
gradually transform from the ones
before the exchanges to the ones after the exchanges. 
We then perform short equilibration simulations without the restraints.
We select the last conformations in the equilibratoin simulations
as ``child'' conformations.
  
In the final stage in Step 2, the selection operation is performed.
We select a superior ``chromosome'' (conformation)
from the parent-child pair.
For this selection operation, 
we employ Metropolis criterion \cite{Metro},
which selects the new child conformation from the parent with 
the following probability:
\begin{equation}
w({\rm p} \rightarrow  {\rm c}) = 
{\rm min} \left( 1, {\rm exp}\{-\beta [ E_{\rm c} - E_{\rm p} ] \}  \right),
\label{eq1}
\end{equation}
where 
$E_{\rm p}$ and $E_{\rm c}$ stand for the potential energy of the parental conformation and 
the child conformation of the parent-child pair, respectively.
$\beta$ is the inverse temperature, which is defined by $\beta = 1/{k_{\rm B} T}$ 
($k_{\rm B}$ is the Boltzmann constant).



\subsection{Combination with replica-exchange method}
The sampling method using genetic crossover in the previous subsection
can be easily combined with other sampling methods such as 
generalized-ensemble algorithms.
Firstly, the conventional MC or MD in Step 1 above can be replaced by
other sampling methods such as MUCA and ST. 
Secondly, the above method can be combined with REM in Step 2 above.

As an example, we introduce a method that combines genetic
crossover and REM.
We first prepare $M$ initial conformations of the system in study, 
where $M$ is the total number of ``individuals'' (in genetic algorithm)
or replicas (in REM)
and is usually taken to be an even integer. While only one temperature
value was used in the previous method, we prepare $M$ different
temperature values $(T_1, \cdots, T_M)$ here. Without loss of generality,
we can assume that $T_1 < \cdots < T_M$.
We then alternately perform the following two steps:
\begin{enumerate}
\renewcommand{\labelenumi}{\arabic{enumi}.}
\item 
For the $M$ individuals, regular canonical MC or MD simulations 
at the fixed temperature $T_m$ $(m=1, \cdots, M)$ 
are carried out simulataneously and independently 
for a certain MC or MD steps.
\item 
$M/2$ pairs of conformations at neighboring temperatures are selected 
from ``parental'' group, and one of the following two operations
is performed.
\begin{enumerate}
\item Two-point genetic crossover is performed for each pair of
parents to produce tow children, and new child conformations are
accepted with the probability in Eq. (\ref{eq1}).
\item each pair of replicas $i$ and $j$ (with coordinates
$q^{[i]}$ and $q^{[j]}$) 
corresponding to neighboring temperatures $T_m$ and $T_{m+1}$,
respectively, is exchanged  with the following probability:
\begin{equation}
w(i \leftrightarrow j) =  {\rm min} \left( 1, {\rm exp}\{-\Delta \}  \right),
\end{equation}
where
\begin{equation}
\Delta = (\beta_m - \beta_{m+1})(E(q^{[j]}) - E(q^{[i]})).
\label{eq2}
\end{equation}
Here, $\beta_m$ is the inverse temperature ($\beta_m = 1/{k_{\rm B} T_m}$)
and $E(q^{[i]})$ is the potential energy of replica $i$ before replica
exchange. If MD is employed in Step 1, we also have to rescale
momenta after replica exchange \cite{REMD}.

\end{enumerate}
\end{enumerate}

If we employ MC simulations in Step 1 above, we can refer the
method to as replica-exchange Monte Carlo using genetic crossover (REMC/GAc)
and if MD simulations, replica-exchange molecular dynamics using genetic 
crossover (REMD/GAc). 
In the above formulation, we chose pairs of
the parent individuals (replicas) that correspond to neighboring 
temperatures. This is to make the acceptance of replica exchange
high. Hence, as far as the crossover operations are concerned,
we could select pairs of parents randomly.

\section{Results and Discussion}
We applied the present methods, namely, PMD/GAc and REMD/GAc,
to Trp-cage.
Trp-cage is known to be one of the smallest protein-like model systems 
and has 20 amino-acid residues. This mini protein was studied 
experimentally by NMR measurements at 282 K (PDB ID: 1L2Y)\cite{Trpcage}.

We incorporated our genetic crossover sampling methods by modifying 
the TINKER program package\cite{tinker_v2}.
The unit time step was set to 1.0 fs.
Each simulation for sampling was carried out for 10.0 nsec (hence, 
it consisted of 10,000,000 MD steps) 
with 16 individuals ($M=16$).
Namely, the total simulation time for sampling was 160.0 nsec.
We performed 200 crossover operations, 
which selected consecutive amino-acid residues of length
between 2 to 10, during the simulations \cite{PSAMDGAc}.
The temperature during MD simulations was controlled by 
Nos\'e-Hoover method \cite{hoover}.
The temperature was set at 282 K for the PMD/GAc simulation 
(the same as the experimental condition).
For REMD/GAc, the number of replicas were also set to $M=16$.
The temperatures were distributed exponentially: 
650, 612, 577, 544, 512, 483, 455, 428, 404, 380, 358, 338, 318, 300, 282, 
and 266 K.
During the REMD/GAc simulation, we performed 100 genetic crossover
operations and 10,000 replica-exchange operations.
As for the conformational potential energy calculations, we used the 
AMBER ff99SB force field \cite{parm99SB}.
As for solvent effects, we used the GB/SA model \cite{gb1,gb2} 
included in the TINKER 
program package \cite{tinker_v2}.
In order to test the effectiveness of the present
method more quantitatively, we have to include more rigorous 
solvation models, which we will do in a future work.
The individuals (replicas) for the simulations had
different sets of randomly generated initial velocities.
We also performed conventional MD and REM simulations for comparisons.
The simulation conditions were the same as above except the crossover 
and selection operations.
In order to balance the computational cost, we performed independent 
16 simulation runs of 10.0 nsec in length 
for the conventional MD.

\begin{figure}
\begin{center}
\resizebox*{8cm}{!}{\includegraphics{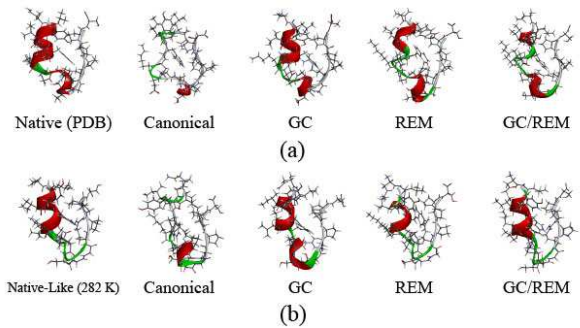}}%
\caption{The lowest-RMSD structures obtained from the 
conventional MD (Canonical), PMD/GAc (GC), conventional
REMD (REM), and REMD/GAc (GC/REM).
(a) is the lowest-RMSD structures with respect to the experimental 
result (Native(PDB), PDB ID:1L2Y model1),
and (b) is the lowest-RMSD structures with respect to the lowest-energy 
conformation obtained from the conventional iso-thermal
canonical simulations at 282 K (Native-Like(282K)).}
\label{fig_min_rmsd_strs}
\end{center}
\end{figure}

In Fig.~\ref{fig_min_rmsd_strs},
we compare the structure of PDB (1L2Y model1) and the lowest-RMSD 
conformations obtained from 
the conventional MD simulation and the PMD/GAc simulation.
The room-mean-square-distance (RMSD) values 
of C$^{\alpha}$ atoms
with respect to the
native structure for the conventional MD simulation and the
PMD/GAc simulation are 4.06 \AA~ and 1.78 \AA, respectively.
Obviously, the conformation obtained from PMD/GAc is more
similar to the native structure than that from 
the conventional MD.
Moreover, we also compared the structures with
a native-like structure, which was the lowest-energy conformation 
obtained from iso-thermal canonical simulations at 282 K.
This native-like conformation was obtained from
16 canonical simulations of 2.0 nsec with different sets of 
randomly generated initial velocities.
The RMSD values with respect to the native-like structure for
the conventional MD simulation and the PMD/GAc simulation
are 3.48 \AA~ and 1.62 \AA, respectively.
The conformation obtained from PMD/GAc is in better agreement with the 
native-like structure than that from the conventional MD.

\begin{figure}
\begin{center}
\resizebox*{8cm}{!}{\includegraphics{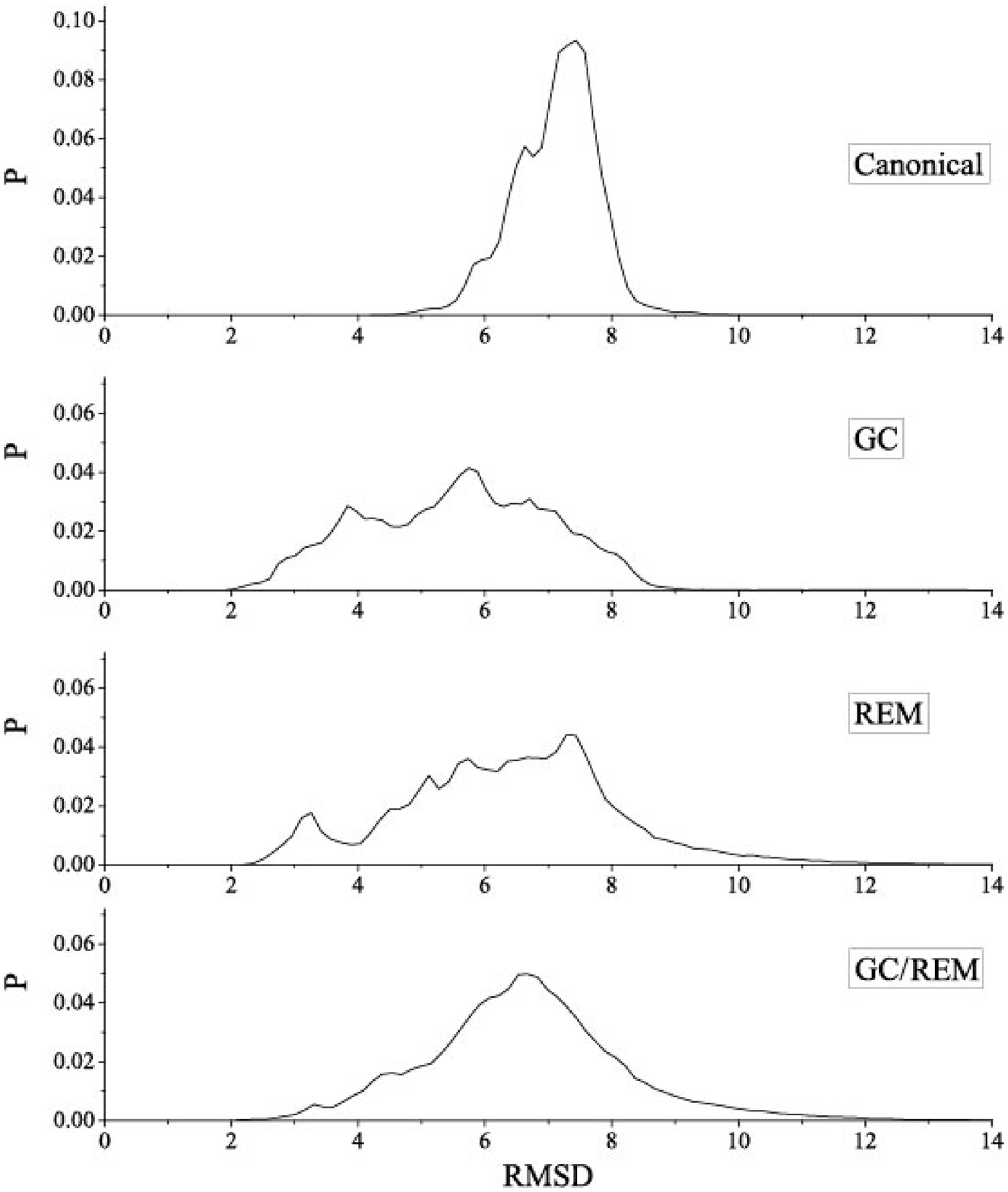}}%
\caption{Probability distributions of RMSD values of all conformations 
at temperature 282 K,} 
obtained from the conventional MD (Canonical), 
PMD/GAc (GC), conventional
REMD (REM), and REMD/GAc (GC/REM).
\label{fig_P_rmsd}
\end{center}
\end{figure}

In Fig.~\ref{fig_P_rmsd}, the probability distributions of RMSD of 
all conformations obtained 
from the conventional MD simulation and the PMD/GAc simulation are shown.
RMSD values obtained from PMD/GAc are lower than those of the 
conventional MD as a whole.
The averages of RMSD values obtained from the conventional MD and 
PMD/GAc are 7.06 \AA~and 5.50 \AA, respectively.
Hence, PMD/GAc can search the conformational space around the 
native structure efficiently  
in comparison with the conventional MD.

We now examine the results of REMD/GAc simulation.
In Fig.~\ref{fig_min_rmsd_strs},  we compare
the lowest-RMSD conformations obtained 
from the conventional REMD simulation and 
REMD/GAc simulation.
The RMSD values with respect to the PDB structure for the conventional 
REMD simulation and the REMD/GAc simulation are 2.03 \AA~ and 1.93 \AA, 
respectively.
Hence, these conformations are almost the same and similar to the PDB structure.
Moreover, the RMSD values with respect to the native-like structure for
the conventional REMD simulation and the REMD/GAc simulation are 
1.78 \AA~ and 1.27 \AA, respectively.
The conformation obtained from REMD/GAc is in slightly better agreement with 
the native-like structure than that from the conventional REMD.

In Fig.~\ref{fig_P_rmsd}, the probability distributions of RMSD of all 
conformations obtained from the conventional REMD simulation and the
REMD/GAc simulation are shown.
The obtained ranges of the RMSD values of both conventional REMD and 
REMD/GAc simulations are broad.
The averages of RMSD values obtained from the conventional REMD simulation
and the REMD/GAc simulations are 6.32 \AA~ and 6.58 \AA, respectively.
Hence, there is almost no difference of the two average values.
However, there are some differences in the distribution of conventional
REMD and REMD/GAc simulations.
The peak values of the probability distributions of the conventional
REMD simulation and the REMD/GAc simulation are 7.28 \AA~ and 6.68 \AA, 
respectively.
Moreover, there is another small peak around 3.26 \AA~ for the conventional
REMD simulation.
On the other hand, there are not any peaks except for
the highest peak in the case of the REMD/GAc simulation.
These results suggest that the REMD/GAc simulation did not get
trapped in any local minima in comparison with the conventional
REMD simulation.

In order to further examine the sampling efficiency of the conventional
REMD simulation and the REMD/GAc simulation, we counted the number of 
tunneling events.
A tunneling event means a random walk from the lowest-energy region to 
the highest-energy region and back, 
and is observed when a system goes from an energy minimum to another 
minimum via a high-energy region \cite{MUCA2,MUCAREM2}.
If the number of the tunneling events is large, the conformational sampling 
is considered to be more efficient \cite{MUCA2,MUCAREM2}.
The numbers of the tunneling events obtained from the conventional
REMD simulation and the REMD/GAc simulation are 54 and 107, respectively 
(these numbers are the total for all the 16 replicas).
We see that REMD/GAc can perform more efficient conformational search 
by using the crossover operation.
Here, the average of the acceptance ratio of crossover 
operations was 0.26. However, the ratio must depend on the system and 
the length $n$ of the crossover operations.


\section{Conclusions}
\label{conclusions}
In this work, we introduced two conformational sampling methods
based on genetic crossover and applied them to a mini protein, Trp-cage.
One method is a combination of conventional molecular dynamics
and genetic crossover, and the other is a further combination
with the replica-exchange method.
These methods realize a broader conformational search by the 
genetic crossover, which is based on global conformational updates.
Conformations close to the native structure were successfully
obtained by these methods.

The genetic crossover sampling methods have a big advantage of 
being highly parallelizable on  parallel computers.
In the future, we are going to apply these methods
to various large proteins in explicit solvent.

\section*{Acknowledgements}
The computations were performed on the computers at the Research Center 
for Computational Science, 
Institute for Molecular Science and Information Technology Center, 
Nagoya University. 
This work was supported, in part, by the
Grants-in-Aid for Scientific Research (A) (No. 25247071),
for Scientific Research on Innovative Areas 
(\lq\lq Dynamical Ordering \& Integrated
Functions\rq\rq ), 
for the Computational Materials Science Initiative, and 
for High Performance Computing Infrastructure from the Ministry of
Education, Culture, Sports, Science and Technology (MEXT), Japan.












%




%











%







~~\\
~~\\

\noindent
{\bf REFERENCES}


%


~~\\
~~\\
~~\\
~~\\
~~\\
~~\\
~~\\
~~\\
~~\\
~~\\

\newpage




\end{document}